\pdfoutput = 1

\documentclass[letterpaper,12pt]{article}   
\usepackage{osajnl2} 
\usepackage{cite}
\usepackage{amsmath}
\usepackage{graphicx}
\usepackage{url}
\usepackage{subfigure}
\usepackage{color}
\usepackage{floatrow}

\usepackage{multirow}
\newcommand{\eqn}[1]{Eq.(\ref{#1})}

\newcommand{\Eqn}[1]{Equation (\ref{#1})}

\newcommand{\Fig}[1]{Figure~\ref{#1}}

\newcommand{\fig}[1]{Fig.~\ref{#1}}

\newcommand{\tab}[1]{table ~\ref{#1}}

\begin{document}

\title{Nonlinear Wave Equation in Frequency Domain: Accurate Modeling of Ultrafast Interaction in Anisotropic Nonlinear Media}


\author{Hairun Guo$^1$, Xianglong Zeng$^{1,2}$, Binbin Zhou$^1$, Morten Bache$^{1,*}$}

\address{$^1$Ultrafast Nonlinear Optics group, DTU Fotonik, Department of Photonics Engineering,Technical University of Denmark, DK-2800 Kgs. Lyngby, Denmark}
\address{$^*$moba@fotonik.dtu.dk}
\address{$^2$Key Laboratory of Special Fiber Optics and Optical Access Networks, SCIE, Shanghai University, Shanghai 200072, China}
\address{zenglong@shu.edu.cn}

\begin{abstract}
  We interpret the purely spectral forward Maxwell equation with up to 3${^{\rm rd}}$ order induced polarizations for pulse propagation and interactions in quadratic nonlinear crystals. The interpreted equation, also named nonlinear wave equation in frequency domain, includes both quadratic and cubic nonlinearities, delayed Raman effects and anisotropic nonlinearities. The full potential of this wave equation is demonstrated by investigating simulations of solitons generated in the process of ultrafast cascaded second-harmonic generation. We show that a balance in the soliton delay can be achieved due to competition between self-steepening, Raman effects and self-steepening-like effects from cascading originating in the group-velocity mismatch between the pump and second harmonic. We analyze the first-order contributions, and show that this balance can be broken to create fast or slow pulses. Through further simulations we demonstrate few-cycle compressed solitons in extremely short crystals, where spectral phenomena such as blue/red shifting, non-stationary radiation in accordance with the non-local phase matching condition and dispersive-wave generation are observed and marked, which help improving the experimental knowledge of cascading nonlinear soliton pulse compression.
\end{abstract}


\maketitle 

\section{Introduction}
\label{sec-intro}
Cascading nonlinearities are known to be produced accompanied with harmonic generation in the limit of large phase-mismatch. As an example phase-mismatched second-harmonic generation (SHG) gives rise to a well-known cascaded quadratic nonlinearity which is Kerr-like and can be written as a nonlinear refractive index ${n_{\rm casc}^{(2)}}$ \cite{desalvo1992self}, i.e. it induces a nonlinear index change that is proportional to the pump intensity. While the energy conversion to the second-harmonic is weak, a Kerr-like nonlinear phase shift is induced on the fundamental wave (FW), which can be large and its sign can be tuned by the phase-mismatch; importantly a negative-sign self-defocusing nonlinearity is accessible. In this case, soliton pulse compression can be fulfilled by combining the self-defocusing cascading nonlinearity with normal dispersion: this so-called cascading nonlinear soliton pulse compressor operates within the visible and near-infrared region and without a power limit compared with the self-focusing pulse compression. Hence, few-cycle and high energy soliton pulses can be generated \cite{ashihara2002soliton,moses2006soliton,zhou2012ultrafast}.

Nonlinear crystals with low dispersion, such as beta-barium borate (BBO), potassium titanyl phosphate (KTP) and lithium niobate (LN), are good candidates for cascading nonlinear soliton pulse compression as they have a decent 2${^{\rm nd}}$ order nonlinear susceptibility, and large cascading nonlinearities can be achieved through tuning the crystal close to phase-matching by either exploiting the birefringent (Type-I) interaction between the FW and SH \cite{nikogosian2005nonlinear}, or by exploiting quasi-phase-matching (QPM) technology. The goal is to overcome the material self-focusing Kerr nonlinearity of the crystal by the self-defocusing cascading nonlinearity, and it can usually be done if the phase-mismatch is taken low enough. Noncritical (Type-0) interactions, where FW and SH are polarized along the crystal axes is desirable because it exploits the largest 2${^{\rm nd}}$ order tensor component, but since the process is not phase-matchable at all, QPM has historically been used to reduce the residual phase-mismatch and thereby increase the cascading strength \cite{kartalouglu1998femtosecond, sundheimer1993large, ashihara2004optical, arbore1997pulse, phillips2011supercontinuum}. In general these nonlinear crystals also promise a stable and compact pulse compressor as they have a short soliton length and a large energy threshold.

Theoretical analysis and experimental demonstrations have been widely reported on cascading nonlinear soliton pulse compression. An elegant theory is that, in the limit of large phase-mismatch, harmonics are all considered as perturbations on the FW. Therefore, the multiple coupled-wave equations (CWEs) can be reduced to a single one governing the FW which is similar to the nonlinear Schr{\"o}odinger equation (NLSE) \cite{moses2006controllable, ilday2004controllable, menyuk1994solitary}. In the NLS-like equation of SHG process, the cascading quadratic nonlinearity is scaled by both the quadratic nonlinearity squared and the inverse of the phase-mismatch. Its properties are revealed through a non-instantaneous ("nonlocal") cascading response which has a Lorentzian shape in frequency domain. This nonlocal response reveals that there are two different characteristic regimes for the response function, known as stationary and non-stationary regimes and the transition between the two regimes is determined largely by the phase-mismatch parameter and the group-velocity mismatch (GVM) parameter \cite{bache2007nonlocal}. A detailed explanation was made by Bache et al. to help understanding most experiments \cite{bache2007scaling, bache2008limits}. For example, the few-cycle high energy pulse compression in Type-I BBO crystal at 1250 nm reported by Moses and Wise \cite{moses2006soliton}, was conducted in a regime in which the phase-mismatch configuration was in the stationary region so that the cascaded response is broadband \cite{bache2007nonlocal} and allow for few-cycle compressed pulses to form. On the other hand, Ashihara et al. reported the moderate pulse compression also using Type-I BBO crystal but launching a pump at 800 nm \cite{ashihara2002soliton}, which is actually in the non-stationary region \cite{bache2007scaling} and the input pulses contained a low soliton order. In both of their experiments, slow pulses with shock fronts are observed which is  induced by GVM as 1${^{\rm st}}$ order cascading response. In a latter report by Ashihara et al., Type-I periodically poled magnesium-oxide-doped lithium niobate (PPMgOLN) was used to compress the pulse at 1560 nm where the GVM between the FW and the SH is suppressed \cite{ashihara2004optical}. Therefore, there is no GVM-related cascading response and the compressed pulse is symmetrical. Recently, the first Type-0 few-cycle pulse compression experiment was demonstrated by Zhou et al., which used a Type-0 LN bulk crystal pumped at 1300 nm \cite{zhou2012ultrafast}, in which the phase-mismatch is located well into the stationary region, see \fig{Fig-Window}(b). The compression was accomplished within an extremely short length of 1 mm as the pulses contained a high soliton order caused by the high intensity and because the FW group velocity dispersion (GVD) in LN is very large. As LN has a dominant delayed Raman response (caused by a strong IR phonon coupling) which continuously transfers energy from high-frequency (blue) components to low-frequency (red) components, tendencies of early pulse splitting were observed.

In this paper, we investigate theoretically and numerically the self-steepening performance and fast/slow pulse tendencies in cascading nonlinear soliton pulse compression induced by both the nonlocal cascading response and the material Raman effects. We do that by directly solving the purely spectral forward Maxwell equation (FME), and a counterbalance of fast/slow pulses is demonstrated. First, we interpret the FME to 3${^{\rm rd}}$ order induced polarizations to describe electric fields interaction in a birefringent material, and model the electric field of the optical wave by directly solving a group of nonlinear wave equations in frequency domain (NWEFs), in which both the 2${^{\rm nd}}$ order and 3${^{\rm rd}}$ order nonlinear-induced polarizations are included, see Sec.\ref{sec-NWEFs}. From the NWEFs, we present the NLS-like equation which includes the system intrinsic self-steepening, a full edition nonlocal cascading response and the material Raman effects, see Sec.\ref{sec-NLSE}. Then, we give a review on the conditions of cascading nonlinear soliton pulse compression (Sec.\ref{sec-condi}) and based on them we discuss the consequences of employing QPM technology. Self-steepening performance and the counterbalance of fast/slow pulses are highlighted as they are greatly dependent on the phase mismatch parameter, see Sec.\ref{sec-QPM}. In Sec.\ref{sec-simul}, we give realistic and convincing simulations to show the self-steepening performance and fast/slow pulses dominated by both the nonlocal cascading response and the material Raman effects. Since the NWEFs are generalized, typical phenomena such as the self phase modulation (SPM) induced broadening in spectrum; generation of dispersive wave and resonant SH radiation in non-stationary regime are still observed. These phenomena can be marked in spectrum, which helps improving the spectral knowledge of cascading nonlinear soliton pulse compression. Finally, we present our conclusions in Sec.\ref{sec-concl}.

\section{Nonlinear Wave Equations in Frequency Domain (NWEFs)}
\label{sec-NWEFs}
In this part, we will present our master equation which directly deals with the electric field. Starting from the 1+1D wave equation of the electric field and making the slowly varying spectral amplitude approximation (SVSAA) \cite{kolesik2012theory}, we obtain the purely spectral FME \cite{bullough1979solitons, husakou2001supercontinuum, conforti2010nonlinear}:

\begin{equation}
\frac{{\partial {\bf{\tilde E}}}}{{\partial z}} + i{\bf{k}}(\omega ){\bf{\tilde E}} =  - i\frac{{{\omega ^2}{\mu _0}}}{{2{\bf{k}}(\omega )}}{{\bf{\tilde P}}_{\rm NL}}
\label{Eq-FME}
\end{equation}

The equation is adaptable to uniaxial, biaxial and cubic/isotropic materials. The "${\sim }$" mark indicates that both the electric field ${\bf{E}}$ and the nonlinear induced polarization ${{\bf{P}}_{\rm NL}}$ are written in frequency domain (we use the Fourier transform ${F[f(t)] = \int_{ - \infty }^{ + \infty } {f(t){e^{ - i\omega t}}dt}}$ and the forward definition as ${{\textstyle{\partial  \over {\partial z}}}{\bf{\tilde E}} \approx  - i{\bf{k}}(\omega ){\bf{\tilde E}}}$). In a uniaxial crystal, ordinary and extraordinary waves, ${E_o}$ and ${E_e}$, are two separate components of the electric field vector, and the propagation constant,${\bf{k}}(\omega ) = \left( {\omega /c} \right){\left( {1 + \tilde \chi (\omega )} \right)^{1/2}}$, is a function of the linear electric susceptibility ${\tilde \chi (\omega )}$ , which is a diagonal matrix containing only the self-response elements \cite{boyd2003nonlinear}.

Among all types of nonlinear induced polarizations, 2${^{\rm nd}}$ order and 3${^{\rm rd}}$ order nonlinear polarization are the most relevant. Normally, the 2${^{\rm nd}}$ order nonlinear polarization, which appears only in non-centrosymmetric media, is considered instantaneous and, in frequency domain, its component can be written as \cite{conforti2010modelling}:

\begin{equation}
\tilde P_j^{(2)} = {\varepsilon _0}\sum\limits_{{\alpha _1},{\alpha _2}} {\left( {\bar \chi _{j;{\alpha _1}{\alpha _2}}^{(2)}F\left[ {{E_{{\alpha _1}}}{E_{{\alpha _2}}}} \right]} \right)}
\label{Eq-chi2}
\end{equation}
where ${j,{a_1},{a_2} = }$ $o$ or $e$, ${\bar \chi _{j;{\alpha _1}{\alpha _2}}^{(2)}}$ is the effective 2${^{\rm nd}}$ order nonlinear susceptibility (with unit ${\rm m/V}$), which is actually the element of a 3-rank tensor. The 3${^{\rm rd}}$ order nonlinear polarization consists of both the instantaneous cubic response (such as electronic Kerr response) and the delayed vibrational Raman response. Its component can be written as \cite{kolesik2004nonlinear}:

\begin{equation}
\tilde P_j^{(3)} = {\varepsilon _0}\sum\limits_{{\alpha _1},{\alpha _2},{\alpha _3}} {\left\{ {\bar \chi _{j;{\alpha _1}{\alpha _2}{\alpha _3}}^{(3)}\left[ {\left( {1 - {f_{\rm R}}} \right)F\left[ {{E_{{\alpha _1}}}{E_{{\alpha _2}}}{E_{{\alpha _3}}}} \right] + {f_{\rm R}}F\left[ {{F^{ - 1}}\left[ {{{\tilde h}_{\rm R}}(\omega )F\left[ {{E_{{\alpha _1}}}{E_{{\alpha _2}}}} \right]} \right]{{\tilde E}_{{\alpha _3}}}} \right]} \right]} \right\}}
\label{Eq-chi3}
\end{equation}
${\bar \chi _{j;{\alpha _1}{\alpha _2}{\alpha _3}}^{(3)}}$ is the effective 3${^{\rm rd}}$ order nonlinear susceptibility (with unit ${\rm m^2/V^2}$) coming from a 4-rank tensor. ${f_{\rm R}}$ indicates the relative fraction of the material Raman effects, ${{\tilde h_{\rm R}}(\omega )}$ is the normalized Raman frequency response function, typically modeled as a complex Lorentzian. Note that if there is no Raman fraction in the material (which is the case for a noble gas), the 3${^{\rm rd}}$ order nonlinear polarization will be completely instantaneous.

Thus, the FME can be expanded to two equations corresponding to the ordinary and extraordinary wave, named nonlinear wave equations in frequency domain (NWEFs):

\begin{multline}
\frac{{\partial {{\tilde E}_o}}}{{\partial z}} + i{k_o}(\omega ){{\tilde E}_o} =  - i\frac{{{\omega ^2}}}{{2{c^2}{k_o}(\omega )}}\left\{ {\sum\limits_{{\alpha _1},{\alpha _2}} {\left( {\bar \chi _{o;{\alpha _1}{\alpha _2}}^{(2)}F\left[ {{E_{{\alpha _1}}}{E_{{\alpha _2}}}} \right]} \right)} } \right.\\
\left. { + \sum\limits_{{\alpha _1},{\alpha _2},{\alpha _3}} {\left( {\bar \chi _{o;{\alpha _1}{\alpha _2}{\alpha _3}}^{(3)}\left( {\left( {1 - {f_{\rm R}}} \right)F\left[ {{E_{{\alpha _1}}}{E_{{\alpha _2}}}{E_{{\alpha _3}}}} \right] + {f_{\rm R}} \cdot F\left[ {{F^{ - 1}}\left[ {{{\tilde h}_{\rm R}}(\omega )F\left[ {{E_{{\alpha _1}}}{E_{{\alpha _2}}}} \right]} \right]{{\tilde E}_{{\alpha _3}}}} \right]} \right)} \right)} } \right\}
\label{Eq-NWEF-O}
\end{multline}
\begin{multline}
\frac{{\partial {{\tilde E}_e}}}{{\partial z}} + i{k_e}(\omega ){{\tilde E}_e} =  - i\frac{{{\omega ^2}}}{{2{c^2}{k_e}(\omega )}}\left\{ {\sum\limits_{{\alpha _1},{\alpha _2}} {\left( {\bar \chi _{e;{\alpha _1}{\alpha _2}}^{(2)}F\left[ {{E_{{\alpha _1}}}{E_{{\alpha _2}}}} \right]} \right)} } \right.\\
\left. { + \sum\limits_{{\alpha _1},{\alpha _2},{\alpha _3}} {\left( {\bar \chi _{e;{\alpha _1}{\alpha _2}{\alpha _3}}^{(3)}\left( {\left( {1 - {f_{\rm R}}} \right)F\left[ {{E_{{\alpha _1}}}{E_{{\alpha _2}}}{E_{{\alpha _3}}}} \right] + {f_{\rm R}} \cdot F\left[ {{F^{ - 1}}\left[ {{{\tilde h}_{\rm R}}(\omega )F\left[ {{E_{{\alpha _1}}}{E_{{\alpha _2}}}} \right]} \right]{{\tilde E}_{{\alpha _3}}}} \right]} \right)} \right)} } \right\}
\label{Eq-NWEF-E}
\end{multline}

We note that in order to solve the above equations, the contents of propagation constants in negative frequencies should be known. According to the causality and reality of the material response, ${{k_j}(\omega ) = k_j^*( - \omega )}$ shows a property of conjugate symmetry \cite{boyd2003nonlinear}. The contents in negative frequencies are therefore linked to that in positive frequencies. Besides, parameters such as ${k(\omega )}$, ${\bar \chi ^{(2)}}$, ${\bar \chi ^{(3)}}$ are dependent on the rotation angle (${\theta }$, ${\varphi }$) of the crystal reference frame \cite{bache2010type, banks2002high, bache2012the}.

Obviously, NWEFs are quite general in physical concept as well as mathematical expression. They describe the dynamics of the electric field rather than the field's envelope. They were directly solved by Husakou et al. with only instantaneous cubic nonlinearities \cite{husakou2001supercontinuum} and by Conforti et al. with only quadratic nonlinearities \cite{conforti2010modelling}. In the present work they are for the first time extended to 3${^{\rm rd}}$ order induced polarization including both instantaneous cubic nonlinearities and delayed Raman effects. By simply including the anisotropy of the nonlinearities, the different wave-mixing possibilities (3- and 4-wave mixing, including third-harmonic generation and parametric up- and down-conversion) under possible phase-matching conditions (Type-0, -I and -II) are automatically modeled properly in the NWEF model.

\section{NLS-like Equation}
\label{sec-NLSE}
From our NWEFs, an NLS-like equation governing the FW envelope can also be derived. First, we  degrade the NWEFs to the common CWEs by employing more approximations such as slowly-varying envelope approximation (SVEA), assumption of constant refractive index around the carrier frequency and assumption of no overlap between the harmonics. Then, in the strong cascading limit ${\Delta k L \gg 1}$ (${\Delta k = {k_2} - 2{k_1}}$ is the phase-mismatch, subscripts 1 and 2 correspond to the FW and the SH, respectively, and ${L}$ is the interaction length) and in the assumption of an undepleted FW (i.e. only the nonlinear phase shift is accumulated as cascading nonlinearity), the ansatz ${{U_2(z,t)} = {A_2}(t)\exp ( - i\Delta kz)}$ is used and the cross phase modulation (XPM) is neglected. Thus, in the FW reference frame we get the NLS-like equation which is more complete compared with what Ilday et al. \cite{ilday2004controllable}, Menyuk et al. \cite{menyuk1994solitary}, Moses et al. \cite{moses2006controllable}, Valiulis et al. \cite{valiulis2011propagation} and Bache et al. \cite{bache2007nonlocal, bache2007scaling} got because the equation includes a full edition nonlocal cascading response as well as the material Raman effects and intrinsic self-steepening. The dimensionless form \cite{bache2008limits} of the complete NLS-like equation is:

\begin{multline}
\left( {i\frac{\partial }{{\partial \xi }} - {D_1}} \right){U_1} + {\rm sgn}(\Delta k) N_{\rm casc}^2\left( {1 - \frac{i}{{{{\hat \omega }_1}}}\frac{\partial }{{\partial \tau }}} \right)U_1^*(\tau )\left( {1 - \frac{i}{{{{\hat \omega }_2}}}\frac{\partial }{{\partial \tau }}} \right)\int_{ - \infty }^\infty  {d\tau '{h_c}(\tau ')U_1^2(\tau  - \tau ')} \\
- N_{\rm cubic}^2\left( {1 - \frac{i}{{{{\hat \omega }_1}}}\frac{\partial }{{\partial \tau }}} \right)\left[ {\left( {1 - {f_R}} \right){{\left| {{U_1}(\tau )} \right|}^2}{U_1}(\tau ) + {f_R}{U_1}(\tau )\int_{ - \infty }^\infty  {d\tau '{h_R}(\tau '){{\left| {{U_1}(\tau  - \tau ')} \right|}^2}} } \right] = 0
\label{Eq-NLSE}
\end{multline}
where ${\xi = z/L_{\rm D,1}}$ and ${\tau = t/T_{\rm 1,in}}$ are dimensionless propagation axis and temporal delay (in the frame of reference traveling with the FW group velocity), respectively, ${L_{\rm D,1}}$ is the dispersion length of the FW and ${T_{\rm 1,in}}$ is input pulse duration. ${U_1}$ is the dimensionless envelope amplitude, ${{D_1} = \sum\nolimits_{m = 2}^\infty  {{\textstyle{{{{( - i)}^m}k_1^{(m)}({{\hat \omega }_1})} \over {m!}}}{\textstyle{{{L_{\rm D,1}}} \over {{T_{\rm 1,in}}^m}}}{{\left( {{\textstyle{\partial  \over {\partial \tau }}}} \right)}^m}} }$ indicates the dimensionless dispersion operator for the FW, in which ${k_j^{(m)}(\hat \omega _j)}$ is the ${m}$-order derivative of ${k_j}$ at ${\hat \omega _j}$ and ${{\hat \omega _j} = {\omega _j}{T_{\rm 1,in}}}$ is the dimensionless frequencies. ${h_{\rm c}(\tau )}$ is the nonlocal cascading response function whose normalized spectral function is ${{\tilde h_{\rm c}}(\Omega ) = {{\Delta k({\omega _1})} \mathord{\left/
{\vphantom {{\Delta k({\omega _1})} {\Delta {k_{\rm nonlocal}}(\Omega )}}} \right. \kern-\nulldelimiterspace} {\Delta {k_{\rm nonlocal}}(\Omega )}}}$, ${\Delta k_{\rm nonlocal}}$ is called nonlocal phase mismatch which will be discussed later. ${h_{\rm R}(\tau )}$ is the material Raman response function. The dimensionless soliton numbers, ${N_{\rm casc}}$ and ${N_{\rm cubic}}$, are linked to the nonlinear refractive indices, i.e. ${N_{\rm casc}^2 = {L_{\rm D,1}}{I_{\rm in}}\frac{{{\omega _1}}}{c}\left| {{n_{\rm casc}^{(2)}}} \right|}$ and ${N_{\rm cubic}^2 = {L_{\rm D,1}}{I_{\rm in}}\frac{{{\omega _1}}}{c}{n_{\rm cubic}}}$, where ${I_{\rm in}}$ is the peak intensity of the launched pump pulse. ${n_{\rm cubic}}$ is the nonlinear refractive index induced by material cubic nonlinearity which consists of electronic Kerr response ${n_{\rm Kerr,el}} = {(1-f_{\rm R})}{n_{\rm cubic}}$ and vibrational Raman response. Compared with the NWEF, the missing effects in the NLS-like equation is: 1) XPM effects and the XPM-induced compression degradation \cite{bache2007scaling, bache2008limits}; 2) divergent cascading response (known as non-stationary regime discussed later) which causes FW depletion; 3) modulated growth of the cascading nonlinearity, which stems from an amplitude modulation of the SH that is neglected in deriving \eqn{Eq-NLSE} \cite{stegeman1996chi2, bache2012higherorder}.

\section{Conditions of Cascading Nonlinear Soliton Pulse Compression}
\label{sec-condi}
In the limit of large phase-mismatch, cascading nonlinearities are produced through the coupling between the FW and the harmonics. Normally, ${n_{\rm casc}^{(2)}}$ dominates the cascading nonlinearity and, particularly in SHG process, it can be expressed in the form of nonlinear refractive index \cite{desalvo1992self}: (in unit of ${ \rm m^2/W}$)

\begin{equation}
n_{\rm casc}^{(2)} =  - \frac{{2{\omega _1}d_{\rm eff}^2}}{{{\varepsilon _0}{c^2}n_1^2{n_2}\Delta k}}
\label{Eq-ncasc}
\end{equation}
where ${\omega _1}$ is the angular frequency of the FW, ${{d_{\rm eff}} = {\bar \chi ^{(2)}}/2}$, ${n_1}$ and ${n_2}$ are refractive indices of the FW and the SH, respectively. \Eqn{Eq-ncasc} means that an equivalent Kerr nonlinearity is generated by the quadratic nonlinearity, and the negative sign indicates that this nonlinearity shows an effect of self-defocusing with a positive ${\Delta k}$ and vice versa. We remind the corresponding electronic Kerr nonlinearity ${n_{\rm Kerr,el}}$ in the material, if the FW photon energy is sufficiently far away from the bandgap in the material, this nonlinearity is positive (self-focusing) \cite{sheik1991dispersion} and therefore forms a competing nonlinearity. The self-defocusing cascading nonlinear soliton pulse compression is therefore only supported if the Kerr-like cascading nonlinearity can counterbalance and exceed the material Kerr nonlinearity and form an overall effect of self-defocusing in the crystal.

\begin{figure}[htb]
  \centering{
  \includegraphics[width = 0.7 \textwidth]{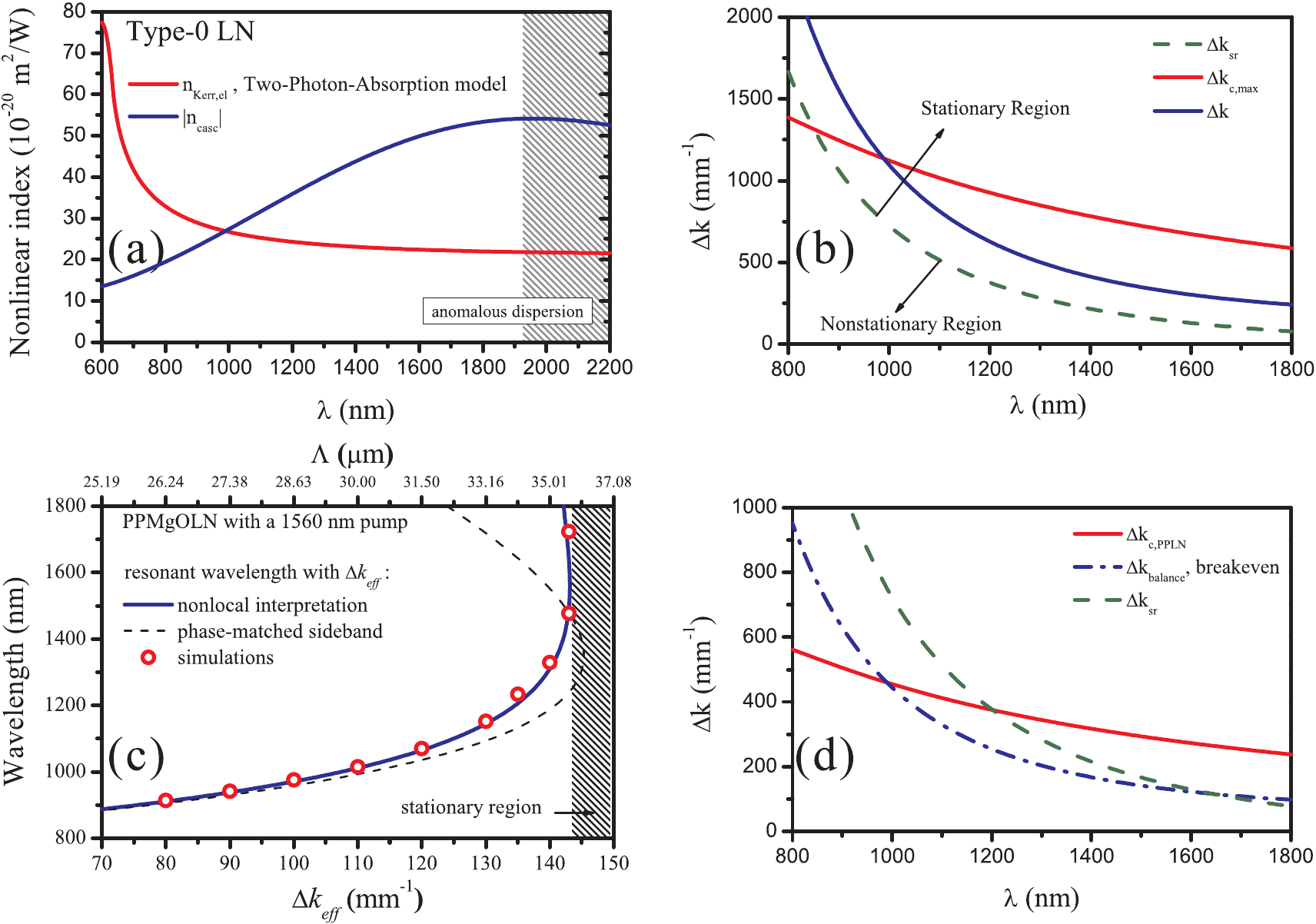}
  }
  \caption{Compression window of MgOLN cut for Type-0 interaction. (a) Cascading quadratic nonlinearity (using ${d_{33} = 25 {\rm pm/V}}$ at 1064 nm and Miller scaling to other wavelengths) and the native electronic Kerr nonlinearity from the so-called two-band model \cite{zhou2012ultrafast, sheik1991dispersion}. (b) Critical boundary of producing an overall self-defocusing nonlinearity and the stationary/nonstationary threshold of the cascading nonlocal response function. (c) Radiation in the nonstationary region, showing the resonant wavelength as a function of the effective phase-mismatch, with the pump located at 1560 nm. (d) Degradation on the compression window if using a 1${^{\rm st}}$ order QPM on MgOLN.}
  \label{Fig-Window}
\end{figure}

Hence, we get the basic condition of our pulse compression, i.e. ${\left| {n_{\rm casc}^{(2)}} \right| > \left| {{n_{\rm Kerr,el}}} \right|}$ , and from this arises a critical boundary of the phase-mismatch, ${\Delta {k_{\rm c}}}$, for which ${n_{\rm casc}^{(2)} + {n_{\rm Kerr,el}} = 0}$. An example is shown in \fig{Fig-Window}(a, b), using MgOLN cut for Type-0 interaction. Within a broad wavelength region (950${\sim }$1920 nm), the self-defocusing ${n_{\rm casc}^{(2)}}$ is stronger than the self-focusing ${n_{\rm Kerr,el}}$  while the dispersion remains normal. In this range an overall effective self-defocusing nonlinearity can be found.

Besides the basic condition of an effective self-defocusing nonlinearity, another condition, determined greatly by the nonlocal phase mismatch ${\Delta {k_{\rm nonlocal}}(\Omega )}$, is necessary if ultrafast interaction is desired as it has a significant influence to the quality of pulse compression. One can namely derive a certain threshold of the phase-mismatch, ${\Delta k_{\rm sr}}$, below which the response of the nonlocal cascading response is resonant (non-stationary), and thus unsuitable for few-cycle pulse generation, while above it is non-resonant and has ultra-broadband (stationary), i.e. ideal for few-cycle interaction. Mathematically, when including complete dispersion the nonlocal phase mismatch can be written as \cite{bache2008limits}:

\begin{equation}
\Delta {k_{\rm nonlocal}}(\Omega ) = k_2\left( {{\omega _2} + \Omega } \right) - 2k_1\left( {{\omega _1}} \right) - \Omega {k_1^{(1)}}\left( {{\omega _1}} \right) = \sum\limits_{m = 2}^\infty  {\frac{{{{\Omega }^m} \cdot {k_2^{(m)}}({\omega _2})}}{{m!}}}  - {d_{12}}\Omega  + \Delta k
\label{Eq-NonPM}
\end{equation}
where ${\omega _2}$ is the frequency of the SH. ${k_j^{(1)}}$ is the group velocity and ${d_{12}}$ is the GVM between the FW and the SH, i.e. ${d_{12}} = {k_1^{(1)}}({\omega _1}) - {k_2^{(1)}}(2{\omega _1})$. Essentially this equation describes the dispersion of the harmonic as observed from pump reference frame and the term ${\Omega {k_1^{(1)}}\left( {{\omega _1}} \right)}$ is physically included to change the moving coordinate system from the the laboratory frame to the FW co-moving frame. A non-resonant cascading response is obtained when ${\Delta {k_{\rm nonlocal}}(\Omega ) \ne 0}$ for any (physical) frequency. The threshold phase-mismatch value ${\Delta k_{\rm sr}}$ marks the boundary where some particular frequency inside the FW spectrum starts to experience nonlocal phase-matching according to this condition. This threshold can be analytically obtained as ${{d_{12}^2}/2k_2^{(2)}}$ if higher order of the SH dispersion is neglected \cite{bache2007nonlocal, bache2007scaling} or like here, be numerically found. In \fig{Fig-Window}(b) we show that in Type-0 MgOLN, the intrinsic ${\Delta k}$ of the material is well within the stationary region. Therefore a clean pulse compression is expected, which can be explained as follows.

Physically, in the non-stationary region, a nonlocal phase-matching condition will be fulfilled between a frequency in the side band of the generated harmonic and the FW center frequency. Due to the narrow bandwidth and large gain of this resonance, a strong oscillation occurs in the temporal profile of the harmonic pulse, resulting in degradation on the compression quality. On the other hand, if ${\Delta k}$ stays above threshold in the stationary region, such a phase-matching condition will not be fulfilled and therefore the harmonic is merely slaved to the FW in the usual manner through ${\tilde A_2 \propto \tilde h_{\rm c}F[U_1^2]}$ (as a part of the cascading channel necessary for generating a FW nonlinear phase shift). In this case the bandwidth of the cascading can be extremely high, even octave-spanning if the phase-mismatch is not too close to the resonance threshold \cite{zhou2012ultrafast}, and in this condition few-cycle soliton formation with a clean temporal profile of the pulse is possible.

When short and broadband pulses interact through (slightly) phase-mismatched harmonic generation, it is common consensus to interpret any resonances in the sidebands of the harmonic spectrum as simply being phase-matching points to a sideband of the FW spectrum, whose frequency detuning is such that the basic harmonic frequency locking is kept, in SHG i.e. ${k_2(\omega_2)\neq 2k_1(\omega_1)}$, but ${k_2(\omega _2 + \Omega) = 2k_1(\omega _1 + \textstyle{\Omega \over 2})}$. However, this phase-matched sideband theory is clearly in contradiction with the nonlocal phase-matching theory here (\eqn{Eq-NonPM}), which dictates that a resonance might occur if a sideband of the SH spectrum can become phase-matched with the FW spectrum taken at its center frequency. We now show that the nonlocal interpretation is the correct one by investigating the spectral behavior of harmonic generation in the non-stationary regime. To do that the phase-mismatch is reduced to below the resonant threshold, which can be achieved in Type-I interaction scheme by employing birefringent angle tuning close to the phase-matching point, or as we do here by employing quasi-phase-matching (QPM) technology in the Type-0 interaction scheme. Several simulations are done with different phase-mismatches and the resonant wavelengths are detected. \Fig{Fig-Window}(c) shows the good agreement between the nonlocal theoretical prediction and the simulation results. Instead the phase-matched sideband theory does not give the correct description except very far from the resonant threshold predicted by the nonlocal theory.

\section{QPM-induced Influence and Self-steepening Performances}
\label{sec-QPM}
QPM technology is well known as a tool to adjust phase-mismatch or even achieve an effective phase-matching. However, the price of using QPM technology is a reduced 2${^{\rm nd}}$ order nonlinear constant due to the pre-factor in the modulation function. For example, ${d_{\rm eff}}$ is a factor ${\textstyle{2 \over \pi }}$ smaller if the 1${^{\rm st}}$ order QPM is employed in the material. A graphical representation of this is that ${\Delta k_{\rm c}}$ is reduced as shown in \fig{Fig-Window}(d), while the benefit is that the phase-mismatch can be flexibly tailored rather than being fixed (in Type-0 interaction scheme). We should therefore rewrite the cascading quadratic nonlinearity in a material with 1${^{\rm st}}$ order QPM structure as:

\begin{equation}
n_{\rm casc,QPM}^{(2)} =  - \frac{{2{\omega _1}{{\left( {{\textstyle{2 \over \pi }}d_{\rm eff}^{}} \right)}^2}}}{{{\varepsilon _0}{c^2}n_1^2{n_2}\Delta {k_{\rm eff}}}}
\label{Eq-ncasc-QPM}
\end{equation}
where ${\Delta {k_{\rm eff}} = \Delta k \pm {\textstyle{{2\pi } \over \Lambda }}}$ and ${\Lambda }$ is the poling period.

Since the reduction of ${d_{\rm eff}}$ is inevitable in the QPM structure, one needs to reduce ${\Delta {k_{\rm eff}}}$ to a level low enough so as to benefit from an increased cascading strength. However, doing this will have some other consequences. 1) The reduced ${\Delta {k_{\rm eff}}}$ will run the risk of getting into the resonant region. As is shown in \fig{Fig-Window}(d), a break-even phase-mismatch value ${\Delta k_{\rm balance}}$ lies where ${n_{\rm casc,QPM}^{(2)} = n_{\rm casc}^{(2)}}$ and it is mostly in the non-stationary region. Therefore in the particular case studied here, Type-0 LN, QPM comes with the price of having a non-stationary cascading nonlinearity except for longer wavelengths close to the zero-dispersion point. 2) The GVM-induced self-steepening effect of cascading will be enhanced as it is scaled by ${\Delta k_{\rm eff}^{-1}}$. It will add to the system intrinsic self-steepening term and therefore give rise to stronger shock front on FW pulses. Moreover, with normal dispersion, self-steepening induced slow pulses will have chance to balance the Raman induced fast pulses.

To analytically prove the latter consequence, we make use of the NLS-like equation. Besides the system's intrinsic self-steepening factor, the self-steepening induced by the cascading is revealed through the 1${^{\rm st}}$ order response. In the weakly nonlocal limit, i.e. ${U_1^2(\tau  - \tau ')}$ is assumed slowly varying compared with the response function. Hence, it can be Taylor expanded as ${U_1^2(\tau  - \tau ') \approx U_1^2(\tau ) - \tau '{\textstyle{\partial  \over {\partial \tau }}}U_1^2(\tau )}$. The nonlocal cascading response (the second term) in \eqn{Eq-NLSE} therefore has:

\begin{multline}
N_{\rm casc}^2\left( {1 - \frac{i}{{{{\hat \omega }_1}}}\frac{\partial }{{\partial \tau }}} \right)U_1^*(\tau )\left( {1 - \frac{i}{{{{\hat \omega }_2}}}\frac{\partial }{{\partial \tau }}} \right)\int_{ - \infty }^\infty  {d\tau '{h_{\rm c}}(\tau ')U_1^2(\tau  - \tau ')} \\
= N_{\rm casc}^2\left( {1 - \frac{i}{{{{\hat \omega }_1}}}\frac{\partial }{{\partial \tau }}} \right)U_1^*\left( {1 - \frac{i}{{{{\hat \omega }_2}}}\frac{\partial }{{\partial \tau }}} \right)\left( {U_1^2 - i{\tau _{\rm c}}\frac{\partial }{{\partial \tau }}U_1^2} \right) \\
= N_{\rm casc}^2\left[ {{{\left| {{U_1}} \right|}^2}{U_1} - \frac{{3i}}{{{\hat \omega _1}}}{{\left| {{U_1}} \right|}^2}\frac{\partial }{{\partial \tau }}{U_1} - 2i{\tau _{\rm c}}{{\left| {{U_1}} \right|}^2}\frac{\partial }{{\partial \tau }}{U_1} - \frac{i}{{{\hat \omega _1}}}U_1^2\frac{\partial }{{\partial \tau }}U_1^*} \right] + HOT
\label{Eq-NLSE-casc}
\end{multline}

where ${{\tau _{\rm c}} = {\tilde h'_{\rm c}}(0) = {d_{12}}/\Delta k({\omega _1}){T_{\rm 1,in}}}$ is defined as the response time of the cascading 1${^{\rm st}}$ perturbation and ${i{\tau _{\rm c}} = \int {d\tau  \cdot \tau  \cdot {h_{\rm c}}(\tau )}}$. It is obvious that a GVM-induced self-steepening term is produced which is scaled by ${\tau _{\rm c}}$ and will directly change the amplitude of the field envelope, resulting in shock front on pulses \cite{moses2006controllable}.

On the other hand, the case of the material Raman effects is quite different. Physically, material Raman effects are well understood to working only on the phase of the field envelope, chirping pulses, transferring energy from the high-frequency components to low-frequency components and resulting in red shifting in the spectrum. Mathematically, if we analogously expand the field intensity ${\left| {{U_1}(\tau  - \tau ')} \right|^2}$ to the 1${^{\rm st}}$ order, the third term in \eqn{Eq-NLSE} has:

\begin{multline}
{\rm{   }}N_{\rm cubic}^2\left( {1 - \frac{i}{{{{\hat \omega }_1}}}\frac{\partial }{{\partial \tau }}} \right)\left[ {\left( {1 - {f_R}} \right){{\left| {{U_1}(\tau )} \right|}^2}{U_1}(\tau ) + {f_R}{U_1}(\tau )\int_{ - \infty }^\infty  {d\tau '{h_{\rm R}}(\tau '){{\left| {{U_1}(\tau  - \tau ')} \right|}^2}} } \right] \\
= N_{\rm cubic}^2\left( {1 - \frac{i}{{{{\hat \omega }_1}}}\frac{\partial }{{\partial \tau }}} \right)\left[ {{{\left| {{U_1}} \right|}^2}{U_1} - {\tau _{\rm R}}{U_1}\frac{\partial }{{\partial \tau }}{{\left| {{U_1}} \right|}^2}} \right] \\
= N_{\rm cubic}^2\left[ {{{\left| {{U_1}} \right|}^2}{U_1} - \left( {\frac{{2i}}{{{\omega _1}}} + {\tau _{\rm R}}} \right){{\left| {{U_1}} \right|}^2}\frac{\partial }{{\partial \tau }}{U_1} - \left( {\frac{i}{{{\omega _1}}} + {\tau _{\rm R}}} \right)U_1^2\frac{\partial }{{\partial \tau }}U_1^*} \right] + HOT
\label{Eq-NLSE-kerr}
\end{multline}
where ${{\tau _{\rm R}} = {f_{\rm R}}\int {d\tau  \cdot \tau  \cdot {h_{\rm R}}(\tau )}}$ is Raman response time. Note that the definitions of ${\tau _{\rm c}}$ and ${\tau _{\rm R}}$ are different as the response time should be physically real-valued. Then, a real part is produced as the 1${^{\rm st}}$ order of the material Raman effects which directly works on the phase of the field envelope.

Furthermore, the field envelope can be separated into a real-valued amplitude multiplying a phase term \cite{ilday2004controllable}, i.e. ${{U_1} = A(\xi ,\tau ){e^{i\varphi (\xi ,\tau )}}}$. In the dispersionless approximation, where the propagation length is much smaller than the dispersion length, the change on the amplitude and phase can be approximately obtained as:

\begin{equation}
\frac{{\partial A}}{{\partial \xi }} = \left( {\frac{{4N_{\rm casc}^2 - 3N_{\rm cubic}^2}}{{{{\hat \omega }_1}}} + 2{\tau _{\rm c}}N_{\rm casc}^2} \right){A^2}\frac{{\partial A}}{{\partial \tau }}
\label{Eq-Amp}
\end{equation}

\begin{equation}
\frac{{\partial \varphi }}{{\partial \xi }} = N_{\rm eff}^2{A^2} - 2{\tau _{\rm R}}N_{\rm cubic}^2A\frac{{\partial A}}{{\partial \tau }} + \left( {\frac{{2N_{\rm casc}^2 - N_{\rm cubic}^2}}{{{{\hat \omega }_1}}} + 2{\tau _{\rm c}}N_{\rm casc}^2} \right){A^2}\frac{{\partial \varphi }}{{\partial \tau }}
\label{Eq-Phase}
\end{equation}
where the effective soliton number, ${{N_{\rm eff}} = \sqrt {N_{\rm casc}^2 - N_{\rm cubic}^2}}$, is defined to quantify the overall self-defocusing nonlinearity. It is obvious that the cascading response time gives contributions to both the amplitude and phase change while the Raman response time only influences the phase change.

\begin{figure}[htbp]
  \centering{
  \includegraphics[width = 0.7 \textwidth]{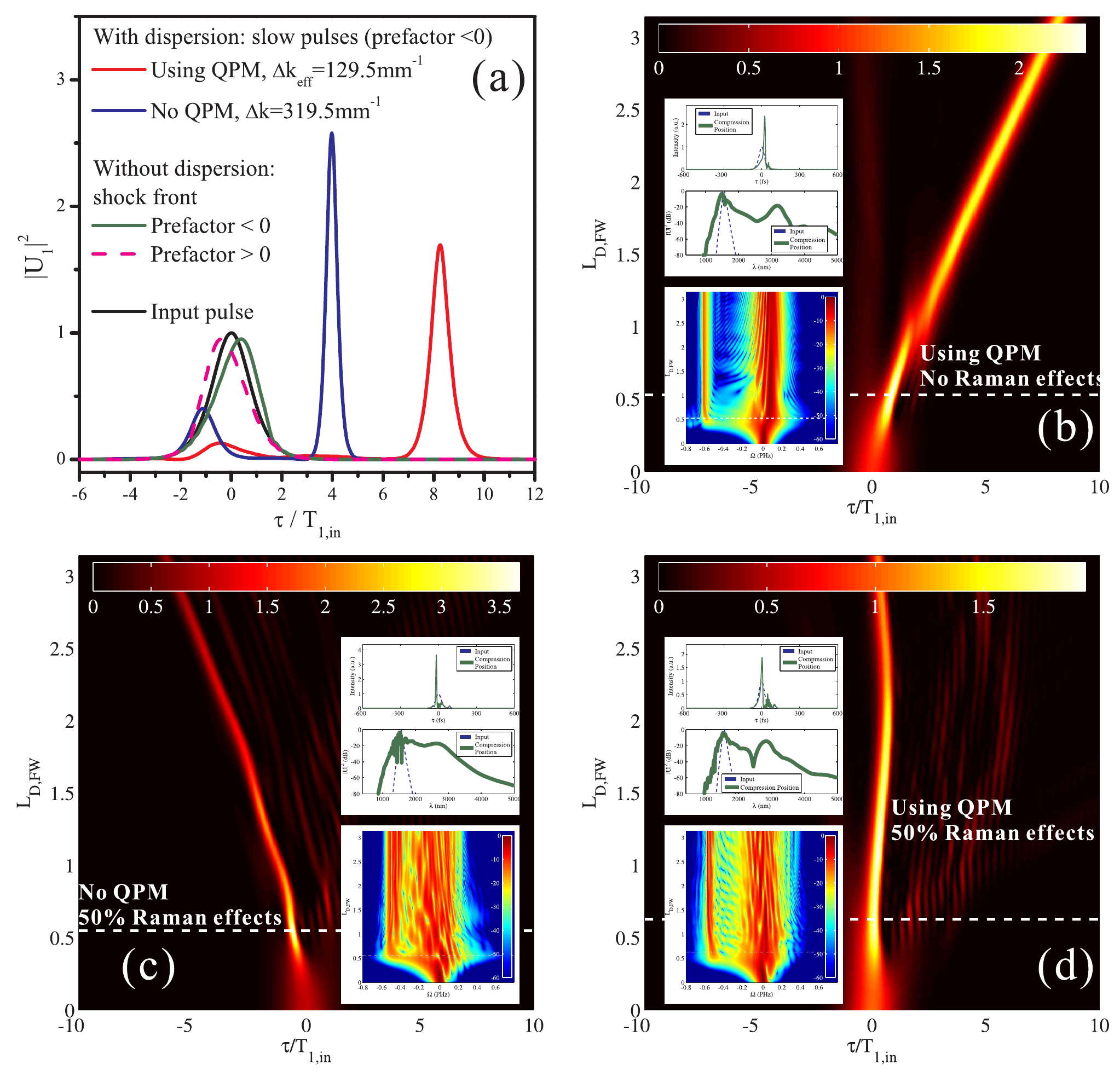}
  }
  \caption{Simulations of NLS-like equation on the balance of the fast/slow pulses. Dispersion properties are chosen from a LN crystal. The cascading nonlinearity is ${N_{\rm casc} = 2.60}$ and the material electronic Kerr nonlinearity is ${N_{\rm Kerr,el} = 1.96}$, the input pulse has a FWHM = 50 fs and a peak intensity ${I_{\rm peak} = 100 {\rm GW/cm^2}}$, located at 1560 nm. (a) shock front on pulses and slow pulses driven by dispersion. (b) slow pulses in QPM crystal without Raman effects, in which ${\Delta k_{\rm eff} = 129.5 {\rm mm^{-1}}}$. (c) fast pulses driven by 50\% Raman effects in a bulk crystal with ${\Delta k = 319.5 {\rm mm^{-1}}}$. (d) balanced pulses generated in a QPM crystal with 50\% Raman effects. The inset figures correspond to the spectrum evolutions, the temporal profiles and the spectra at the first compression point of the pulses (position marked by the white dash line).}
  \label{Fig-NLSE}
\end{figure}

The self-steepening induced shock front on pulses is therefore understood to happen when the pre-factor on the right part of the amplitude equation (\eqn{Eq-Amp}) is not zero, see \fig{Fig-NLSE}(a), in which ${\tau _{\rm c}}$ and ${N_{\rm casc}}$ play a role as tuning parameters. A similar conclusion was reported by Moses and Wise \cite{moses2006controllable} but they did not get further into the amplitude and phase dynamics. Moreover, the shock front also causes asymmetry in pulse spectrum and, in the existence of the material dispersion, therefore generates fast/slow pulses. \fig{Fig-NLSE}(a) also shows that for lower (effective) phase mismatch, slower pulses are generated.

For the phase equation (\eqn{Eq-Phase}), we note that if a stable soliton is formed during the propagation, the phase chirp becomes negligible (the third term can be eliminated as ${{\textstyle{{\partial \varphi } \over {\partial t}}} = 0}$ for a stable soliton) and the Raman term turns to dominate (the second term), which continuously gives rise to red shifting in pulse spectrum. Then, with normal dispersion, fast pulses are generated due to the faster GV at the red-shifted wavelengths.

Hence, we could expect a balance of the fast/slow pulses with the competing interaction of the cascading effects and the material Raman effects. Convincing numerical results are shown in \fig{Fig-NLSE}(b-d) through direct simulations of \eqn{Eq-NLSE}, in which the cascading nonlinearity and the material electronic Kerr nonlinearity are always kept identical. Without Raman effects, the cascading response time will be tuned by phase mismatch parameter and generates slow pulses, see \fig{Fig-NLSE}(b); When the Raman term is switched on as 50\% of the total cubic nonlinearity, fast pulses are driven as a counterbalance to the cascading effects. In a bulk crystal where the cascading response time is quite short due to the large ${\Delta k}$, eventually fast pulses are generated, see \fig{Fig-NLSE}(c). When using a QPM crystal, however, the cascading response is enhanced and may form a balance with the Raman effects, which keeps the pulses around the zero delay position, see \fig{Fig-NLSE}(d). The oscillations split out of the pulses are dispersive waves (D-wave, also called Cherenkov wave), which are the result of a phase-matching condition of the linear part of the spectrum  (the anomalous dispersion regime at longer wavelengths) to the soliton residing in the normal dispersion regime.

\section{Simulations and Discussions}
\label{sec-simul}
In this part, realistic simulation examples based on our master equations are shown. Few-cycle soliton pulses are generated within a short crystal length from multi-cycle pump pulses. Such a compressor has therefore great potential for improving the performance of femtosecond laser systems.

The samples we take into consideration are BBO, MgOLN and PPMgOLN, which are all uniaxial crystals within the point group ${3m}$. If these crystals are operated far from the band gap resonance, they are assumed to be under the Kleinman symmetry. Thus there are only a few non-zero elements in the nonlinear tensors, e.g. only 3 independent elements in 2${^{\rm nd}}$ order nonlinear tensor and 4 in the 3${^{\rm rd}}$ order nonlinear tensor. The nonlinear parameters used in the simulations here are listed in \tab{tab-para}. The cubic nonlinear parameters of BBO were taken from a recent literature study \cite{bache2012the}, while those for MgOLN were taken from \cite{zhou2012ultrafast}. Note that the large Raman fraction of LN was found there by comparing simulations with experimental data, as no reliable absolute Raman nonlinear coefficients could be found in the literature.

\begin{table}[htb]

  \begin{tabular}{ | l | c | c | c | c |}
  \hline
  \multicolumn{2}{|c|}{Material} & BBO (1.03${\rm \mu m}$) & MgOLN (1.3${\rm \mu m}$) & PPMgOLN (1.3${\rm \mu m}$) \\ \hline
  \multicolumn{1}{|l|}{\multirow{3}{0.22 \textwidth}
  {2${^{\rm nd}}$ order nonlinear tensor elements (${\rm pm/V}$)}}
  & ${d_{22}}$ & 2.20 \cite{nikogosian2005nonlinear} & /      & /      \\
  & ${d_{31}}$ &-0.04 \cite{nikogosian2005nonlinear} & /      & /      \\
  & ${d_{33}}$ &-0.04 \cite{nikogosian2005nonlinear} & -23.50 & -23.50 \\
  \hline
  \multicolumn{1}{|l|}{\multirow{5}{0.22 \textwidth}
  {3${^{\rm rd}}$ order nonlinear tensor elements (${\rm pm^2/V^2}$)}}
  & ${c_{11}}   $ &  550.00 \cite{bache2012the}& /    & /    \\
  & ${c_{33}}   $ &-1400.00 \cite{bache2012the}& 7300 & 7300 \\
  & ${c_{16}}   $ &  120.00 \cite{bache2012the}& /    & /    \\
  & ${c_{10}}   $ &  -22.00 \cite{bache2012the}& /    & /    \\
  & ${f_{\rm R}}$ &  0                         & 50\% & 50\% \\
  \hline
  \multicolumn{2}{|l|}{\multirow{2}{0.25 \textwidth}
  {Interaction type:}}
  & Type-I, Type-II & Type-0 & Type-0, QPM           \\
  \cline{3-5}
    \multicolumn{1}{|c}{}
  & \multicolumn{1}{c|}{}
  & \includegraphics[width = 0.2 \textwidth]{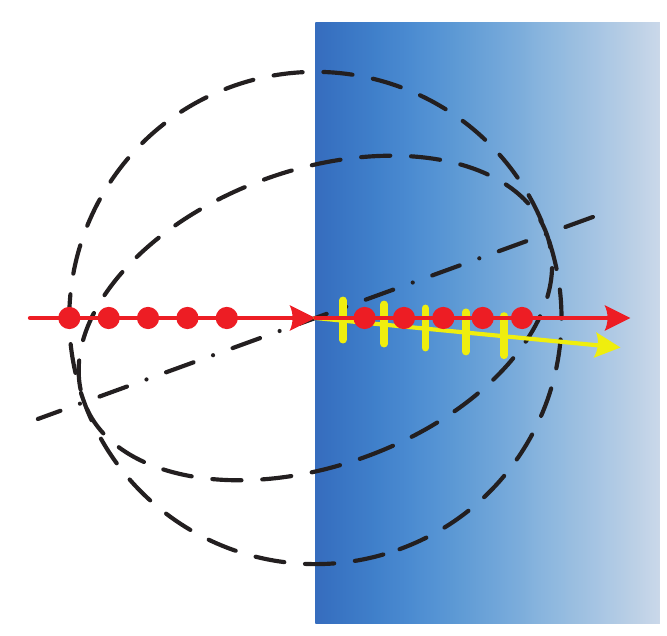}
  & \includegraphics[width = 0.2 \textwidth]{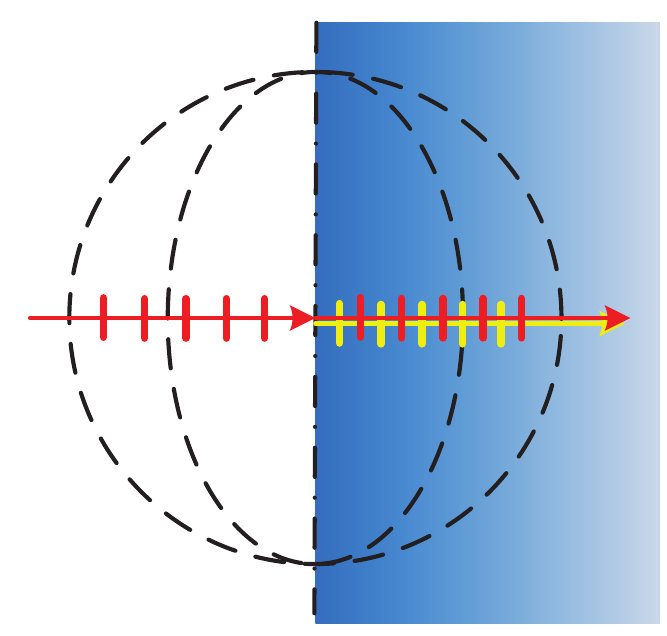}
  & \includegraphics[width = 0.2 \textwidth]{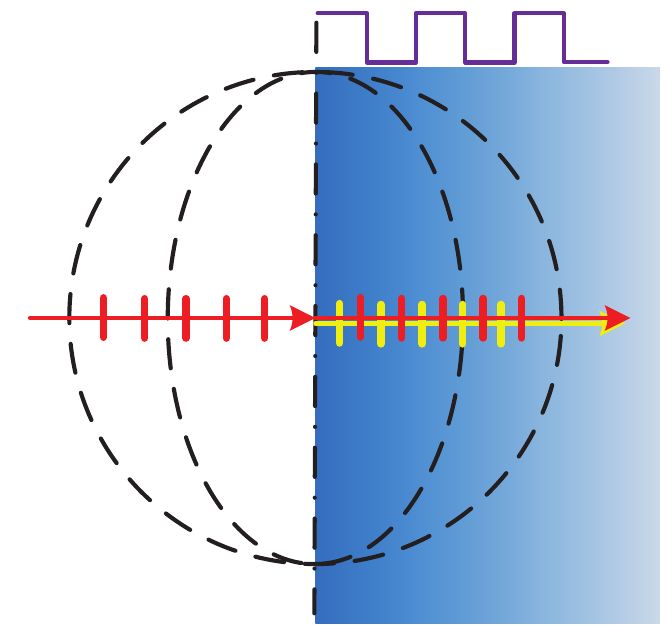} \\
  \hline
  \end{tabular}
  \caption{Nonlinear parameters used for simulations}
  \label{tab-para}
\end{table}

\subsection{Type-I BBO at 1030 nm: A Clean Few-Cycle Soliton Pulse Compressor}
\begin{figure}[htb]
  \centering{
  \includegraphics[width = 0.7 \textwidth]{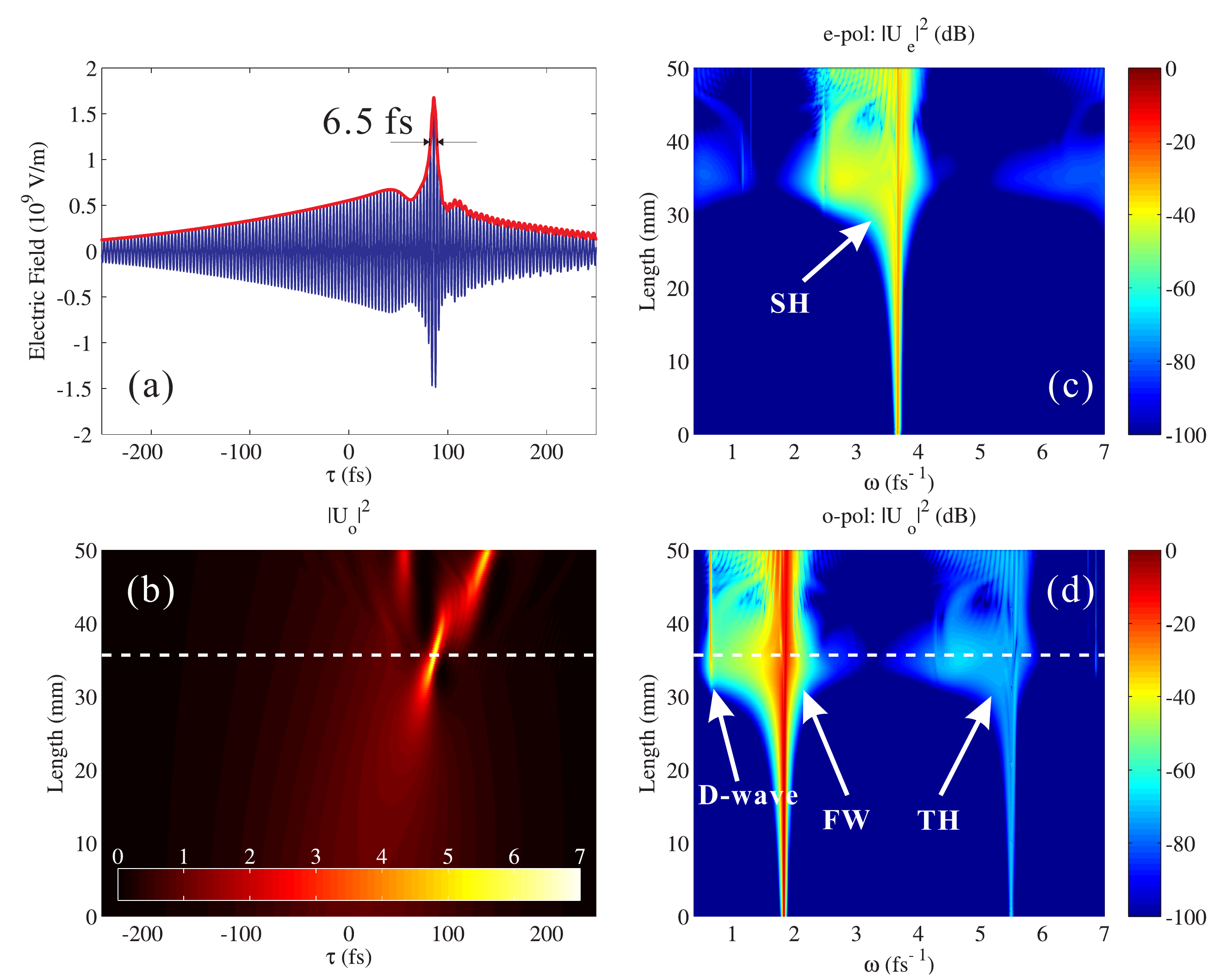}
  }
  \caption{Simulation of cascading nonlinear soliton pulse compression in BBO cut for Type-I interaction at 1030 nm. The rotation (${\theta }$, ${\varphi }$) = (20.5${^0}$, -90.0${^0}$), under which ${\Delta k_{\rm c} = 78.1 {\rm mm^{-1}}}$ and ${\Delta k_{\rm sr} = 46.1 {\rm mm^{-1}}}$. Native phase-mismatch ${\Delta k = 55.4 {\rm mm^{-1}}}$. The launched pulse has FWHM = 200 fs, peak intensity ${I_{\rm in} = 100 {\rm GW/cm^2}}$, ${n_{\rm casc}^{(2)} = -7.86\times 10^{-20} {\rm m^2/W}}$, ${n_{\rm Kerr,el|o:ooo} = 5.67\times 10^{-20} {\rm m^2/W}}$, ${N_{\rm eff} = 6.00}$. (a) The electric field of the 1${^{\rm st}}$ stage soliton (position marked by the white dash line); (b) The temporal evolution of the FW; (c, d) Spectral evolutions of both o-pol and e-pol pulses.}
  \label{Fig-NWEF-BBO}
\end{figure}

A BBO crystal cut for Type-I interaction has a compression window (950-1450 nm \cite{bache2012the}, where the two compression conditions are both satisfied) for longer near-IR wavelengths \cite{bache2007nonlocal}, which can be accessed by tuning the crystal rotation angle ${\theta }$. Unfortunately, the most concerned Ti:sapphire laser wavelength, 800 nm, is not included in such a window and therefore only moderate compression under a non-stationary regime could be achieved \cite{ashihara2002soliton}. Within this window, clean pulse compression is expected. In fact the successful generation of few-cycle compressed pulses at 1260 nm by Moses and Wise was done at such conditions. In \fig{Fig-NWEF-BBO} we investigate a similar scenario at 1030nm where high energy sub-picosecond pulses can be generated by an ytterbium-doped laser system. While the phonon response of BBO has been measured \cite{ney1998assignment}, its strength is generally believed to be very weak, so in the simulations we take ${f_{\rm R} = 0}$. Moreover, it is relevant to mention that BBO has normal FW dispersion below 1450 nm.

The electric field of the compressed pulse is shown in \fig{Fig-NWEF-BBO}(a). After few centimeters propagation, clean and sub-3-cycle pulses are generated. Besides, the cascading response causes shock front on the pulses and generate slow pulses through normal dispersion, see \fig{Fig-NWEF-BBO}(b) showing the temporal evolution. In the spectral evolution \fig{Fig-NWEF-BBO}(c, d), the FW and several harmonics are marked and cascading nonlinearities are produced among them. A weak SH is generated through the phase-mismatched SHG process (Type-I, o+o:e) while a weaker third harmonic (TH) is generated through the phase-mismatched sum frequency generation (SFG) process (Type-II, o+e:o) as well as the phase-mismatched third harmonic generation (THG) process (o+o+o:o). Since the SHG process has the smallest phase-mismatch (${\rm \sim 55 mm^{-1}}$) and moderate conversion efficiency compared with other processes, the cascading quadratic nonlinearity dominates the total self-defocusing nonlinearities.

It is to our knowledge the first time that such purely spectral NWEF(s) are used to model the few-cycle soliton pulse generation based on a total self-defocusing nonlinearity originating from the competition between the cascading nonlinearity and material intrinsic Kerr nonlinearity. Moreover, we performed a comparison between the results of the NWEFs and the usual CWEs and found only a slight difference in the pulse's pedestal; this difference is caused by the THG that is included in the NWEFs, but since the TH is weak here only a slight difference is observed. Thus, the type I simulations in BBO traditionally done with the usual coupled-wave equations based on the SEWA \cite{moses2006soliton, ashihara2002soliton, moses2006controllable, ilday2004controllable, bache2007nonlocal, bache2007scaling, bache2008limits} turns out to give a fairly accurate description of the dynamics in cascaded soliton compression. Nonetheless, the advantage of the NWEFs is that it automatically takes into account any type of multistep wave mixing and harmonic conversion, which cannot a priori be excluded.

\subsection{Type-0 MgOLN: Few-Cycle, Fast Pulses Dominated by the Material Raman Effects}
\begin{figure}[htb]
  \centering{
  \includegraphics[width = 0.7 \textwidth]{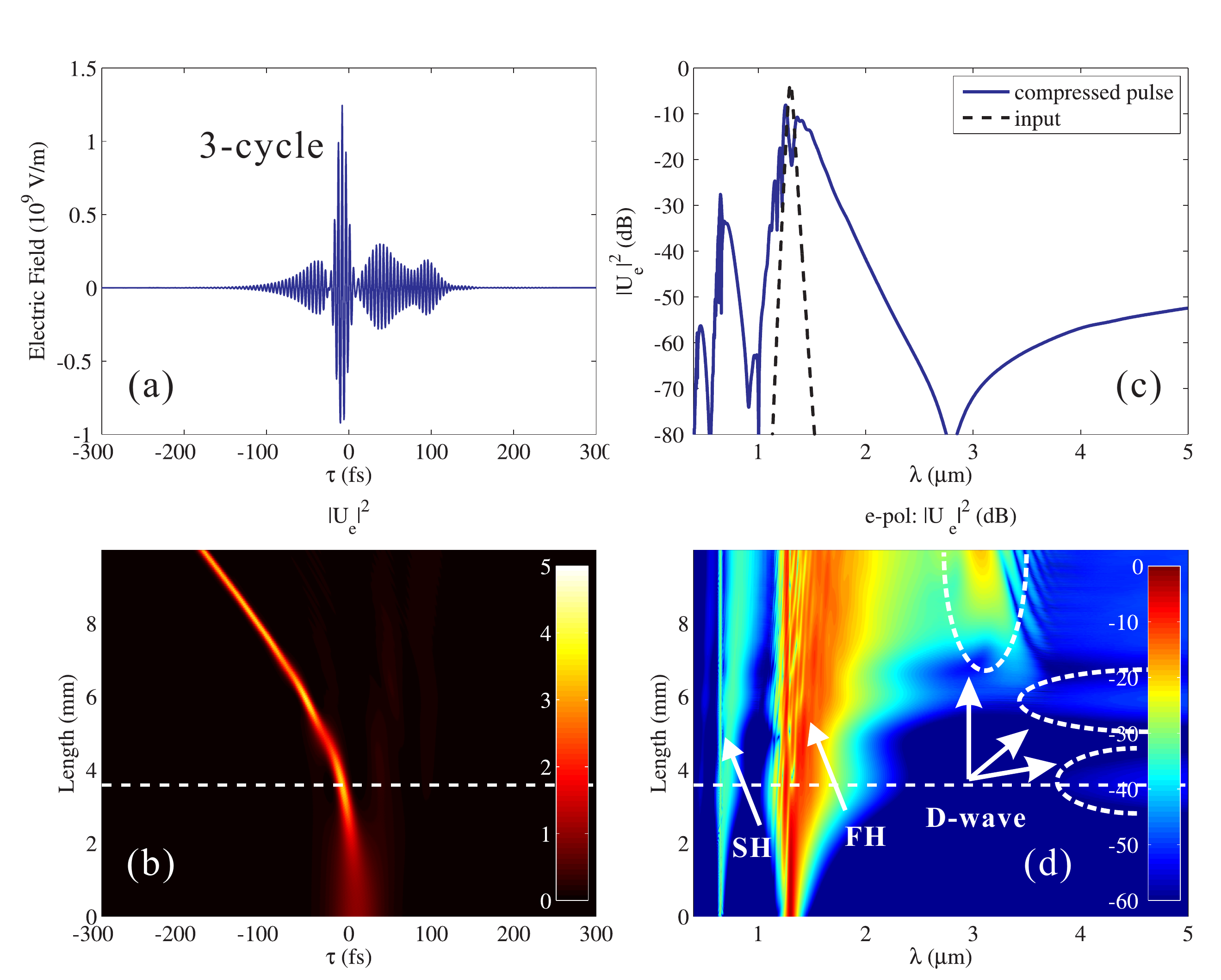}
  }
  \caption{Simulation of cascading nonlinear soliton pulse compression in MgOLN cut for Type-0 interaction at 1300 nm. The rotation (${\theta }$, ${\varphi }$) = (90${^0}$, 90${^0}$), i.e. X-cut, under which ${\Delta k_{\rm c} = 890.4 {\rm mm^{-1}}}$ and ${\Delta k_{\rm sr} = 283.3 {\rm  mm^{-1}}}$. Native phase-mismatch ${\Delta k = 501.5 {\rm mm^{-1}}}$. The launched pulse has FWHM = 50 fs, peak intensity ${I_{\rm in} = 100 {\rm GW/cm^2}}$, ${n_{\rm casc}^{(2)} = -40.05\times 10^{-20} {\rm m^2/W}}$, ${n_{\rm Kerr,el} = 22.56\times 10^{-20} {\rm m^2/W}}$, ${\sqrt {N_{\rm casc}^2 - N_{\rm Kerr,el}^2} = 2.05}$. (a) The electric field of the 1${^{\rm st}}$ stage soliton (position marked by the white dash line); (b) The temporal evolution of the FW; (c) The spectrum of the 1${^{\rm st}}$ stage soliton; (d) The spectral evolution of the e-pol pulse with obvious red shifting of the FW and blue shifting of the D-wave.}
  \label{Fig-NWEF-MgOLN}
\end{figure}

MgOLN cut for Type-0 interaction is also a good candidate for cascading nonlinear pulse compression due to its large ${d_{33}}$ element. The compression window has already been shown in \fig{Fig-Window}. However, in MgOLN, it seems that the Raman effect is quite dominant \cite{zhou2012ultrafast}, giving a large relative fraction ${f_{\rm R}}$, which causes fast pulses in time domain and red-shifting in spectrum, see \fig{Fig-NWEF-MgOLN}.

By launching a short 50 fs pump, 3-cycle pulses are generated within a short propagation length of around ${\rm 4 mm}$ (\fig{Fig-NWEF-MgOLN}(a, b)) because MgOLN has a large GVD. It is highlighted that fast pulses are dramatically driven after the soliton formation in which the Raman chirping term starts to dominate the phase dynamics (\eqn{Eq-Phase}). Correspondingly, obvious red shifting in spectral evolution is observed and the D-wave becomes more blue shifted, see \fig{Fig-NWEF-MgOLN}(d), a consequence of the phase-matching law between them. Compared with Zhou's experiment \cite{zhou2012ultrafast}, the pulse splitting is not prominent here as we use a lower soliton order and therefore reduce the effect of Raman-induced soliton fission.

\subsection{Type-0 PPMgOLN: Few-cycle, Zero Delay Pulses Generation with Enhanced D-wave}

\begin{figure}[htb]
  \centering{
  \includegraphics[width = 0.7 \textwidth]{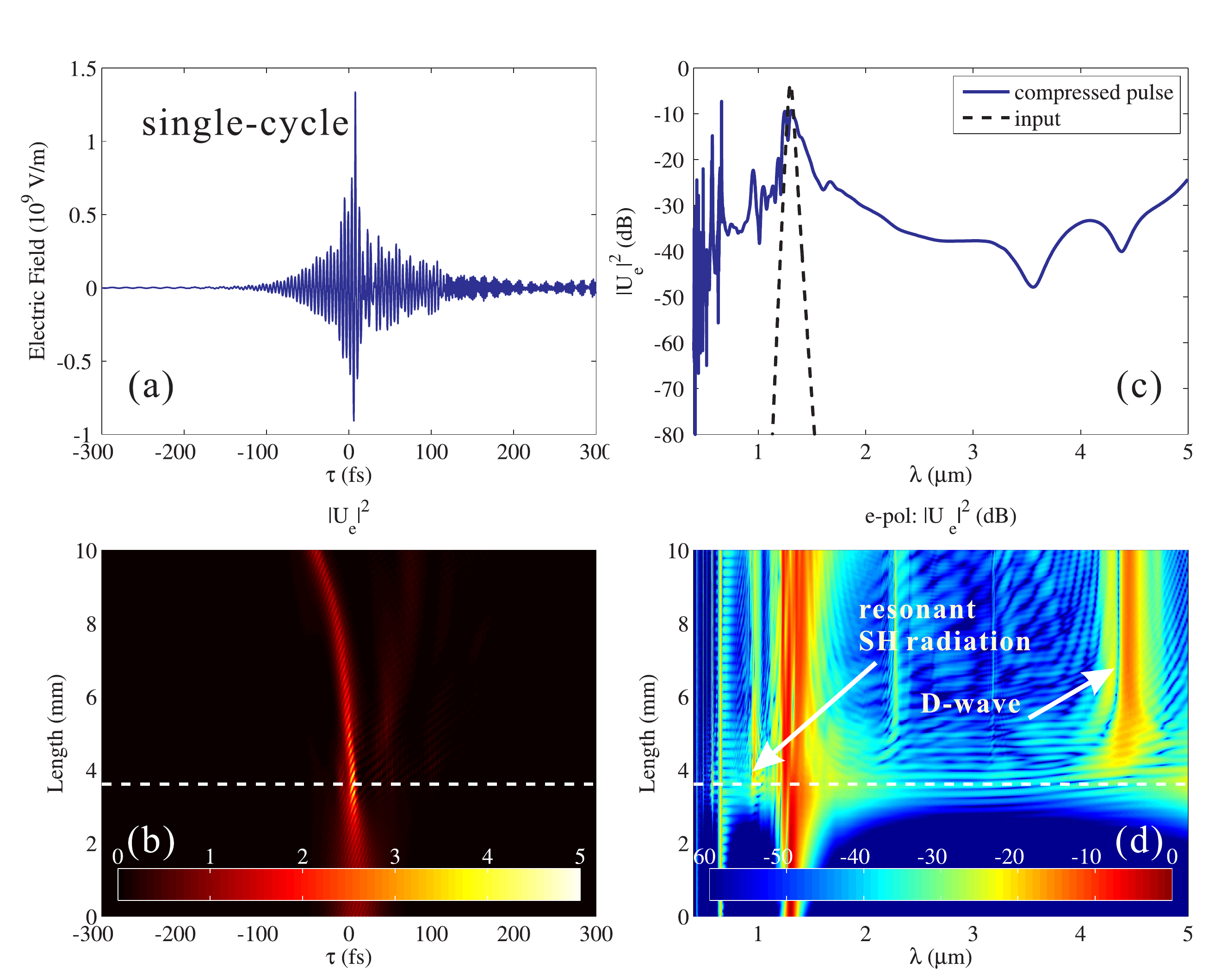}
  }
  \caption{Simulation of cascading nonlinear soliton pulse compression in PPMgOLN cut for Type-0 interaction at 1300 nm. The poling period is 26 ${\rm \mu m}$. The rotation (${\theta }$, ${\varphi }$) = (90${^0}$, 90${^0}$), under which ${\Delta k_{\rm c,QPM} = 360.9 {\rm  mm^{-1}}}$ and ${\Delta k_{\rm sr} = 283.3 {\rm  mm^{-1}}}$. The effective phase-mismatch ${\Delta k_{\rm eff} = 259.8 {\rm mm^{-1}}}$. The launched pulse has FWHM = 50 fs, peak intensity ${I_{\rm in} = 100 {\rm GW/cm^2}}$, ${n_{\rm casc,QPM}^{(2)} = -31.32\times 10^{-20} {\rm m^2/W}}$, ${n_{\rm Kerr,el} = 22.56\times 10^{-20} {\rm m^2/W}}$, ${\sqrt {N_{\rm casc,QPM}^2 - N_{\rm Kerr,el}^2} = 1.45}$. (a) The electric field of the pulse (position marked by the white dash line); (b) The temporal evolution of the FW; (c) The spectrum of the pulse; (d) The spectral evolution of the pulse with resonant SH radiation and strong dispersive wave.}
  \label{Fig-NWEF-PPLN}
\end{figure}

To make a comparison with the bulk MgOLN crystal, pulse compression in a PPMgOLN with an effective phase-mismatch in the non-stationary region is shown in \fig{Fig-NWEF-PPLN}. Using the same pump, the compressed soliton has a single-cycle spike, but overall the soliton is not as clean as \fig{Fig-NWEF-PPLN} and has a stronger shock front. With a great reduction in the phase-mismatch, from ${501.5 {\rm mm^{-1}}}$ to ${259.8 {\rm mm^{-1}}}$ through a 26 ${\mu {\rm m}}$ QPM pitch, the 1${^{\rm st}}$ order response of the cascading (scaled by ${\tau _{\rm c}}$) is enhanced which pulls the pulses back to the zero delay position, see \fig{Fig-NWEF-PPLN}(b) showing the temporal evolution. In the spectral evolution \fig{Fig-NWEF-PPLN}(d), Raman induced red shifting is well suppressed and the D-wave is evoked and enhanced around ${4.5 {\rm \mu m}}$. Besides, the self-steepening pre-factor is also enlarged, resulting the strong shock front on the pulses, see \fig{Fig-NWEF-PPLN}(a). Meanwhile, in the non-stationary region, nonlocal phase matching is fulfilled so that a strong resonant peak emerges in the spectrum, see \fig{Fig-NWEF-PPLN}(c,d), which generates strong temporal oscillations and degrades the compression quality.

\section{Conclusions}
\label{sec-concl}
As a conclusion, in this paper we presented the generalized nonlinear wave equations in frequency domain (NWEFs), which are for the first time extended to 3${^{\rm rd}}$ order induced polarization including both instantaneous cubic nonlinearities and delayed Raman effects. Any of the possible phase-matching conditions (Type-0, -I and -II) are easily modeled, and by including the anisotropy of the nonlinearities, the different wave-mixing possibilities (3- and 4-wave mixing, including third-harmonic generation and parametric up- and down-conversion) are automatically included properly. We then used the NWEFs as a platform for investigating ultrafast cascaded second-harmonic generation, in which soliton compression to few-cycle duration is possible in short nonlinear frequency conversion crystals.

We first reduced the NWEFs to a single NLS-like equation, which highlights self-steepening effects as well as the competition between cascaded nonlocal effects and cubic SPM and Raman effects. Using this reduced equation we discussed the conditions for observing optimal pulse compression.  It is common to interpret the resonant SH peak, which can be observed in cascading when the phase mismatch is small, as phase-matching of the spectral sidebands. We showed that this is wrong and that it can instead be accurately described by the nonlocal theory as a phase-matching between an offset SH frequency with the FW center frequency.

The reduced NLS-like equation was also used to investigate the competition between self-steepening, Kerr self-focusing SPM, and Raman effects on one side, and on the other side cascaded contributions. Besides the self-defocusing SPM term, these cascading terms to first order contribute both to self-steepening-like terms (creating a pulse front shock) and also to higher-order phase contributions similar to the Raman effect. Since the cascading terms are tunable through the sizes and signs of the phase-mismatch and the GVM parameters, we showed that the slow/fast pulses usually associated with self-steepening and Raman effects could be balanced out by cascading: The Raman red-shift of the pulse gave fast solitons due to the faster GV at red-shifted wavelengths (normal dispersion), while GVM-induced cascading tends to slow pulses down. Therefore, pulses where the two delay effects are balanced are expected and we showed simulations where these effects appeared. In fact, we even found cases where the strong spectral red-shift from Raman was completely suppressed by strong cascading contributions.

Finally, we showed a series of full simulations of the NWEFs, where few-cycle, clean soliton pulse compression was demonstrated. In BBO the Type-I simulations showed spectral components related to third-harmonic generation and sum-frequency mixing, but we found that the results agreed well with previous simulations of the coupled SEWA equations. In MgOLN a strong Raman contribution was investigated in the Type-0 configuration, and we showed that in a bulk crystal a few-cycle soliton forms after a few millimeters, but also that due to the dominating Raman effect the soliton is accelerated. We also performed the same simulation with the same pump but in a QPM (through periodic poling) MgOLN, where the phase mismatch was greatly reduced. In this case the cascading response is no longer optimal as it becomes spectrally resonant. The QPM case gave more characteristic phenomena in the spectrum but the temporal profile was not so clean. In addition, the soliton was no longer accelerated by Raman effects, and spectrally the Raman red-shifting was cancelled. This is a consequence of the increased cascading self-steepening from QPM.

\section*{Acknowledgement}
\label{sec-ackno}

Xianglong Zeng acknowledges the support of Marie Curie International Incoming Fellowship from EU (grant No. PIIF-GA-2009¨C253289) and the financial support from National Natural Science Foundation of China (60978004) and Shanghai Shuguang Program (10SG38).Morten Bache acknowledges the support from the Danish Council for Independent Research (Technology
and Production Sciences, grant No. 274-08-0479 ¡±Femto-VINIR¡±, and grant No. 11-106702 ¡±Femto-midIR¡±).


\end{document}